\documentclass[12pt,onecolumn]{IEEEtran}
\usepackage{amsmath}
\usepackage{amssymb}
\usepackage{cite}
\usepackage{color}
\usepackage{epsfig}
\usepackage{epsf}
\usepackage{rotating}

\renewcommand{\baselinestretch}{1.8}
\newtheorem{thm}{Theorem}
\newtheorem{lemma}{Lemma}

\begin{document}

\thispagestyle{empty} \vskip 1cm

\begin{center}
\vskip 0.6cm
{\large \bf LLL Reduction Achieves the Receive Diversity in MIMO Decoding}\footnote{This work was supported in part by funding from Communications and Information Technology Ontario (CITO), Nortel Networks, and Natural Sciences
and Engineering Research Council of Canada (NSERC). The material of this paper was presented at the IEEE International
Symposium on Information Theory, Adelaide, Australia, September 2005.} \\

\vskip 0.8cm Mahmoud Taherzadeh, Amin Mobasher, and Amir K. Khandani

\vskip 0.5cm

{\small Coding \& Signal Transmission Laboratory\\
Department of Electrical \& Computer Engineering\\
University of Waterloo\\
Waterloo, Ontario, Canada, N2L 3G1\\
}

\end{center}

\begin{abstract}
Diversity order is an important measure for the performance of communication systems over MIMO fading channels. In this paper, we prove that in MIMO multiple access systems (or MIMO point-to-point systems with V-BLAST transmission), lattice-reduction-aided decoding achieves the maximum receive diversity (which is equal to the number of receive antennas). Also, we prove that the naive lattice decoding (which discards the out-of-region decoded points) achieves the maximum diversity.

\end{abstract}

\section{Introduction}

In the recent years, MIMO communications over multiple-antenna
channels has attracted the attention of many researchers. In \cite{Fos96}, a transmission technique called V-BLAST is introduced for high-rate communications over point-to-point MIMO fading channels. V-BLAST sends independent symbols over different transmit antennas. Therefore, it can also be used for MIMO multi-access systems. Most of the sub-optimum decoding methods for BLAST (such as nulling and cancelling, zero forcing and GDFE-type methods) can not achieve the maximum receive diversity which is equal to the number of receive antennas. In \cite{damen2000}, a lattice decoder is proposed for the decoding of BLAST which (according to the simulation results) achieves the maximum diversity. However, its complexity is exponential in terms of the number of antennas. In \cite{fischer2003}, \cite{Wornell}, and \cite{Mow}, an approximation of lattice decoding, using the LLL lattice-basis reduction \cite{LLL82}, is introduced which has a polynomial complexity and the simulation results show that it achieves the receive diversity. In this paper, we give a mathematical proof for achieving the receive diversity by the LLL-aided zero-forcing decoder, which is one of the simplest forms of the lattice-reduction-aided decoders. Also, a similar proof shows that the naive lattice decoding (which discards the out-of-region decoded points) achieves the receive diversity.

\section{Basic concepts and System Model}

A real (or complex) lattice $ \Lambda $ is a discrete set
of $ N $-dimensional vectors in the real Euclidean space $ \mathbb{R}^{N} $
(or the complex Euclidean space $ \mathbb{C}^{N} $) that forms a
group under ordinary vector addition. Every lattice $ \Lambda $ is
generated by the linear combinations of a set of linearly
independent vectors $ \mathbf{b}_{1},\cdots ,\mathbf{b}_{M} \in
\Lambda $, with integer (or Gaussian integer) coefficients.
The set of vectors $ \left\lbrace \mathbf{b}_{1},\cdots ,\mathbf{b}_{M} \right\rbrace $ is
called a basis of $ \Lambda $, and the $N \times M $ matrix $ \mathbf{B} = \left[ \mathbf{b}_{1},
\cdots , \mathbf{b}_{M}\right]  $, which has the basis vectors as its
columns, is called the generator matrix of $
\Lambda $.

The basis of the lattice is not unique. Indeed, we can obtain a new generator matrix by multiplying the old generator matrix by any $M\times M $ unimodular matrix, where a unimodular matrix is defined as an integer matrix whose inverse has also integer entries. In many applications, a basis consisting of relatively short and nearly orthogonal
vectors is desirable. The procedure of finding such a basis for a
lattice is called \textit{Lattice Basis Reduction}. In \cite{LLL82}, a basis-reduction algorithm, the so-called LLL basis reduction, is introdiced which results in relatively
short basis vectors with a polynomial-time computational complexity. 

We consider a multiple-antenna system with $M$ transmit antennas and $N$ receive antennas, where $M \leq N $. 
In a multiple-access system, we
consider different transmit antennas as different users. We
consider vectors $ \mathbf{y}=[y_{1},...,y_{N}]^{T} $, $
\mathbf{x}=[x_{1},...,x_{M}]^{T} $, $
\mathbf{w}=[w_{1},...,w_{N}]^{T} $ and the $ N\times M $ matrix
$\mathbf{H}$, as the received signal, the transmitted
signal, the noise vector and the channel matrix, respectively\footnote{In this paper, boldface small letters represent vectors; boldface capital letters represent matrices, $(\cdot)^{\mathtt{H}} $ denotes the Hermitian of a matrix and $(\cdot)^{-\mathtt{H}}$ denotes $ {\left( (\cdot)^{\mathtt{H}}\right)}^{-1} $.}. The
following matrix equation describes the channel model:
\begin{equation}
\mathbf{y}=\mathbf{H}\mathbf{x}+\mathbf{w} .
\end{equation}
The channel is assumed to be Raleigh and the noise is Gaussian, i.e.
the elements of $\mathbf{H}$ are i.i.d with the zero-mean
unit-variance complex Gaussian distribution. Also, we have the power
constraint on the transmitted signal, $ \textmd{E} \Vert \mathbf{x} \Vert^{2}
=1$. The power of the additive noise is $ \sigma^{2} $ per antenna,
i.e. $ \textmd{E} \Vert \mathbf{w} \Vert^{2} =N\sigma^{2} $. Therefore, the
signal to noise ratio (SNR) is defined as $
\rho=\frac{1}{\sigma^{2}} $.

In a MIMO multiple-access system or a MIMO point-to-point system with V-BLAST transmission, we send the transmitted vector $\mathbf{x}$ with independent entries from $ \mathbb{Z}[i] $, the set of complex Gaussian integers. At the receiver, as the maximum-likelihood (ML) estimate of $\mathbf{x} $, a vector $\widehat{\mathbf{x}} $ should be found among the possible transmitted vectors, such that $\Vert\mathbf{y} -\mathbf{H}\widehat{\mathbf{x}} \Vert$ is minimized. For large constellations, the exact ML decoding can be very complex and practically infeasible. Therefore, we need to approximate it by a low-complexity scheme. 

As a simple approximation of ML decoding, zero-forcing can be used, which selects $\widehat{\mathbf{x}} $ as the closest integer point to $\mathbf{H}^{-1}\mathbf{y} $. Although zero forcing is very simple to implement, it has a poor performance. Indeed, in zero forcing, $\mathbf{H}^{-1}\mathbf{w} $ is the effective noise, and when $\mathbf{H} $ has a small singular value, $\mathbf{H}^{-1}$ can have very large row vectors, which result in magnifying the effective noise power. To overcome this shortcoming of the zero-forcing decoder, lattice-basis reduction is used in \cite{fischer2003}, \cite{Wornell}, and \cite{Mow} to enhance the performance of zero forcing and reduce its effective noise.

We can perform two slightly different types of LLL-aided decoding:

\textbf{Type I)} We find $ \tilde{\mathbf{x}} $ as the closest integer point to $ \mathbf{B}^{\mathtt{H} }\mathbf{y} $ where the $N\times M $ matrix $ \mathbf{B} $ is the reduced version of $ \mathbf{H}^{-\mathtt{H} } $, i.e. $ \mathbf{B}=\mathbf{H}^{-\mathtt{H} } \mathbf{U} $, where $ \mathbf{U} $ is an $M\times M $ unimodular matrix (when $ M<N$, we use the pseudo-inverse instead of the inverse). The transmitted vector is decoded as,
$$ \widehat{\mathbf{x}}=\mathbf{U}^{-\mathtt{H} }\tilde{\mathbf{x}}. $$
In the absence of noise (when $\mathbf{w}=\mathbf{0}$), 
$$\widehat{\mathbf{x}}=\mathbf{U}^{-\mathtt{H} }\tilde{\mathbf{x}} =\mathbf{U}^{-\mathtt{H} }\mathbf{B}^{\mathtt{H} }\mathbf{y}=\mathbf{U}^{-\mathtt{H} }\left(\mathbf{H}^{-\mathtt{H} } \mathbf{U} \right) ^{\mathtt{H} }\mathbf{y}=\mathbf{U}^{-\mathtt{H} }\mathbf{U}^{\mathtt{H} }\mathbf{H}^{-1} \mathbf{H}\mathbf{x}=\mathbf{x}.$$
In the presence of the noise, $ \mathbf{B}^{\mathtt{H} }\mathbf{w}$ can be seen as the effective noise (instead of $ \mathbf{H}^{-1}\mathbf{w}$ in the traditional zero forcing).

\textbf{Type II)} We find $ \tilde{\mathbf{x}} $ as the closest integer point to $ {\mathbf{H}_{red}}^{-1}\mathbf{y} $ where $ \mathbf{H}_{red} $ is the reduced version of $ \mathbf{H}$ i.e. $ \mathbf{H}_{red}=\mathbf{H} \mathbf{U} $. The transmitted vector is decoded as,
$$ \widehat{\mathbf{x}}=\mathbf{U}\tilde{\mathbf{x}} . $$
In the absence of noise (when $\mathbf{w}=\mathbf{0}$), 
$$\widehat{\mathbf{x}}=\mathbf{U}\tilde{\mathbf{x}}=\mathbf{U}{\mathbf{H}_{red}}^{-1}\mathbf{y}=\mathbf{U}\mathbf{U}^{-1} \mathbf{H}^{-1}\mathbf{H}\mathbf{x}=\mathbf{x}$$
In the presence of the noise, $ {\mathbf{H}_{red}}^{-1}\mathbf{w}$ is the effective noise.

In the previous works \cite{fischer2003} \cite{Wornell} \cite{Mow}, the LLL-aided decoding type II has been used. We show that the type I method is more appropriate to reduce the effective noise, and indeed, has a better performance. In the next section, we present the details of the proof of our main result for the first method and show that a similar proof is valid for the second method. 

\section{Diversity of LLL-aided decoding}

For MIMO systems, diversity is defined as $\lim_{\rho\rightarrow \infty} \dfrac{-\log P_{e}}{\log \rho} $. When there is no joint processing among the transmit antennas, the maximum achievable diversity is equal to $N$, the number of receive antennas \cite{Tarokh98}. To prove that LLL-aided decoding achieves a diversity order of $N$, we use a bound on $ \delta$, the orthogonality defect of the LLL reduction, which is defined as 
$$\delta = \frac{(\Vert \mathbf{b}_{1}\Vert^{2} \Vert \mathbf{b}_{2}\Vert^{2}...\Vert
\mathbf{b}_{M}\Vert^{2})}{\det \mathbf{B^{\mathtt{H} }B}}.
$$

\begin{thm} [see \cite{Napias}]
Let $\Lambda$ be an $ M $-dimensional complex lattice and $\mathbf{ B} =[\mathbf{b}_{1}...\mathbf{b}_{M}]   $ be the LLL reduced basis of $
\Lambda $. If $ \delta$ is the orthogonality defect of $ \mathbf{B}$, then,
\begin{equation} \sqrt{\delta} \leq 2^{M(M-1)} .
\end{equation}
\end{thm}

In the rest of this section, in the lemmas 1-3, we bound the error probability by the probability of an inequality on $ d_{\mathbf{H}}$ (the minimum distance among the points of the lattice generated by $ \mathbf{H}$) and the length of the noise vector being valid. In lemma 4, we bound the probability that $ d_{\mathbf{H}}$ is too small. Finally, in theorem 2, we prove the main result by combining the bounds on the probability that $ d_{\mathbf{H}}$ is too small, and the probability that the noise vector is too large.

\begin{lemma}
Consider $ \mathbf{B}=[\mathbf{b}_{1}...\mathbf{b}_{M}] $ as an $ N
\times M $ matrix, with the orthogonality defect $ \delta $,  and $
\mathbf{B}^{-\mathtt{H} }= [\mathbf{a}_{1}...\mathbf{a}_{M}]$ as the
Hermitian of its inverse (or its pseudo-inverse if $M<N$). Then\footnote{This lemma is an extention of lemma 1 in \cite{TechRep}.},
\begin{equation} \label{eq:lemma1-1}
\max\lbrace \Vert \mathbf{b}_{1}\Vert,...,\Vert \mathbf{b}_{M}\Vert \rbrace \leq \frac{\sqrt{\delta}}{\min\lbrace \Vert \mathbf{a}_{1}\Vert,...,\Vert \mathbf{a}_{M}\Vert \rbrace} 
\end{equation}
and
\begin{equation} \label{eq:lemma1-2}
\max\lbrace \Vert \mathbf{a}_{1}\Vert,...,\Vert \mathbf{a}_{M}\Vert \rbrace \leq \frac{\sqrt{\delta}}{\min\lbrace \Vert \mathbf{b}_{1}\Vert,...,\Vert \mathbf{b}_{M}\Vert \rbrace} .
\end{equation}
\end{lemma}

\begin{proof} Consider $ \mathbf{b}_{i} $ as an arbitrary column of $
\mathbf{B} $. The vector $ \mathbf{b}_{i} $ can be written as $
\mathbf{b}_{i}^{\prime} + \sum_{i \neq j} c_{i,j} \mathbf{b}_{j} $,
where $ \mathbf{b}_{i}^{\prime} $ is orthogonal to $ \mathbf{b}_{j}
$ for $ i \neq j $. Now, $ [\mathbf{b}_{1}...\mathbf{b}_{i-1} \mathbf{b}_{i}^{\prime} \mathbf{b}_{i+1}...\mathbf{b}_{M}]$ can be written as $ \mathbf{BP}$ where $ \mathbf{P}$ is a unit-determinant $ M \times M$ matrix (a column operation matrix):
\begin{equation} \Vert \mathbf{b}_{1} \Vert^{2}...\Vert\mathbf{b}_{i-1}\Vert^{2} . \Vert \mathbf{b}_{i} \Vert^{2} . \Vert \mathbf{b}_{i+1}\Vert^{2}...\Vert \mathbf{b}_{M} \Vert^{2} \end{equation}
\begin{equation}
= \delta \det \mathbf{B^{\mathtt{H} }B} = \delta  \det \mathbf{P}^{\mathtt{H} }\mathbf{B^{\mathtt{H} }}  \mathbf{B} \mathbf{P} 
\end{equation}
 \begin{equation} = \delta \det\left(  [\mathbf{b}_{1}...\mathbf{b}_{i-1} \mathbf{b}_{i}^{\prime} \mathbf{b}_{i+1}...\mathbf{b}_{M}]^{\mathtt{H} }[\mathbf{b}_{1}...\mathbf{b}_{i-1} \mathbf{b}_{i}^{\prime} \mathbf{b}_{i+1}...\mathbf{b}_{M}]\right) .  \end{equation}
According to the Hadamard theorem:
\begin{equation}
\det\left(  [\mathbf{b}_{1}...\mathbf{b}_{i-1} \mathbf{b}_{i}^{\prime} \mathbf{b}_{i+1}...\mathbf{b}_{M}]^{\mathtt{H} }[\mathbf{b}_{1}...\mathbf{b}_{i-1} \mathbf{b}_{i}^{\prime} \mathbf{b}_{i+1}...\mathbf{b}_{M}]\right)  \leq
 \end{equation}
 \begin{equation}
 \Vert \mathbf{b}_{1} \Vert^{2}...\Vert\mathbf{b}_{i-1}\Vert^{2} . \Vert \mathbf{b}_{i}^{\prime} \Vert^{2} . \Vert \mathbf{b}_{i+1}\Vert^{2}...\Vert \mathbf{b}_{M} \Vert^{2}.
\end{equation}
Therefore,
\begin{equation}
\Vert \mathbf{b}_{1} \Vert^{2}...\Vert\mathbf{b}_{i-1}\Vert^{2} . \Vert \mathbf{b}_{i} \Vert^{2} . \Vert \mathbf{b}_{i+1}\Vert^{2}...\Vert \mathbf{b}_{M} \Vert^{2}\leq   \delta \Vert \mathbf{b}_{1} \Vert^{2}...\Vert\mathbf{b}_{i-1}\Vert^{2} . \Vert \mathbf{b}_{i}^{\prime} \Vert^{2} . \Vert \mathbf{b}_{i+1}\Vert^{2}...\Vert \mathbf{b}_{M} \Vert^{2}  
\end{equation}
\begin{equation}
\Longrightarrow  \Vert \mathbf{b}_{i} \Vert \leq  \sqrt{\delta} \Vert \mathbf{b}_{i}^{\prime} \Vert.
\end{equation}
Also, $ \mathbf{B}^{-1}\mathbf{B}=\mathbf{I}$ results in
$<\!\!\mathbf{a}_{i},\mathbf{b}_{i}\!\!>\;= 1$ and
$<\!\!\mathbf{a}_{i},\mathbf{b}_{j}\!\!>\;= 0 $ for $ i\neq j $.
Therefore,
\begin{equation}  1 \; =\;\;<\!\!\mathbf{a}_{i},\mathbf{b}_{i}\!\!>\;\;= \;\; <\!\!\mathbf{a}_{i},(\mathbf{b}_{i}^{\prime} + \sum_{i \neq j} c_{i,j} \mathbf{b}_{j})\!\!> \;\;= \;\; <\!\!\mathbf{a}_{i},\mathbf{b}_{i}^{\prime}\!\!> \;\; 
\end{equation}
Now, $\mathbf{a}_{i}$ and $\mathbf{b}_{i}^{\prime} $, both are orthogonal to the $(M-1)$-dimensional subspace generated by the vectors $\mathbf{b}_{j} $ ($ j\neq i$). Thus, 
\begin{equation}
1 = \;\; <\!\!\mathbf{a}_{i},\mathbf{b}_{i}^{\prime}\!\!> \;\; \; = \; \Vert\mathbf{a}_{i}\Vert .\Vert\mathbf{b}_{i}^{\prime}\Vert \geq \Vert\mathbf{a}_{i}\Vert . \frac{\Vert\mathbf{b}_{i}\Vert}{\sqrt{\delta}}\end{equation}
\begin{equation} \Longrightarrow 1 \geq \Vert\mathbf{b}_{i}\Vert . \frac{\Vert\mathbf{a}_{i}\Vert}{\sqrt{\delta}} \end{equation}
\begin{equation} \label{eq:15}  \Longrightarrow \Vert\mathbf{b}_{i}\Vert \leq \frac{\sqrt{\delta}}{\Vert\mathbf{a}_{i}\Vert} \end{equation}
The above relation is valid for every $ i $, $ 1\leq i\leq M $.
Without loss of generality, we can assume that $
\max\lbrace \Vert \mathbf{b}_{1}\Vert,...,\Vert \mathbf{b}_{M}\Vert
\rbrace= \Vert \mathbf{b}_{k} \Vert $:
\begin{equation} \max\lbrace \Vert \mathbf{b}_{1}\Vert,...,\Vert \mathbf{b}_{M}\Vert \rbrace= \Vert \mathbf{b}_{k} \Vert \leq \frac{\sqrt{\delta}}{\Vert\mathbf{a}_{k}\Vert} \end{equation}
  \begin{equation}\leq \frac{\sqrt{\delta}}{\min\lbrace \Vert \mathbf{a}_{1}\Vert,...,\Vert \mathbf{a}_{M}\Vert \rbrace} . \end{equation}
    Similarly, by using (\ref{eq:15}), we can also obtain the following inequality:
 \begin{equation} 
\max\lbrace \Vert \mathbf{a}_{1}\Vert,...,\Vert \mathbf{a}_{M}\Vert \rbrace \leq \frac{\sqrt{\delta}}{\min\lbrace \Vert \mathbf{b}_{1}\Vert,...,\Vert \mathbf{b}_{M}\Vert \rbrace} .
\end{equation} 
\end{proof}

\begin{lemma}
Consider $ \mathbf{B}=[\mathbf{b}_{1}...\mathbf{b}_{M}] $ as a reduced basis (LLL) \cite{LLL82} for the lattice generated by $ \mathbf{H}^{-\mathtt{H} } $, $ \mathbf{B}^{-\mathtt{H} } = [\mathbf{a}_{1} ...\mathbf{a}_{M}] $, and $ \delta $ as the orthogonality defect of the reduction. Then, if the magnitude of the noise vector is less than $ \dfrac{ \min\lbrace \Vert \mathbf{a}_{1}\Vert,...,\Vert \mathbf{a}_{M}\Vert \rbrace}{2\sqrt{M \delta}}$, the LLL-aided decoding method correctly decodes the transmitted signal.
\end{lemma}

\begin{proof} When we use the LLL-aided decoding method, we find the nearest integer point to $ \mathbf{B}^\mathtt{H}\mathbf{y} $. We should show that this point is the same as the transmitted vector; or in other words, all the elements of $ \mathbf{B}^\mathtt{H}\mathbf{w} $ are in the interval $ (-\frac{1}{2},\frac{1}{2}) $. To prove this, we show that $ \Vert \mathbf{B}^\mathtt{H}\mathbf{w} \Vert < \frac{1}{2} $. It is easy to show that,

\begin{equation} \Vert \mathbf{B}^\mathtt{H}\mathbf{w} \Vert \leq \sqrt{M}\cdot  \max\lbrace \Vert \mathbf{b}_{1}\Vert,...,\Vert \mathbf{b}_{M}\Vert \rbrace \cdot \Vert \mathbf{w} \Vert 
\end{equation}

Now, according to (\ref{eq:lemma1-1}),

\begin{equation} \max\lbrace \Vert \mathbf{b}_{1}\Vert,...,\Vert \mathbf{b}_{M}\Vert \rbrace \leq \frac{\sqrt{\delta}}{\min\lbrace \Vert \mathbf{a}_{1}\Vert,...,\Vert \mathbf{a}_{M}\Vert \rbrace} 
\end{equation}
Therefore,

\begin{equation} \Vert \mathbf{B}^\mathtt{H}\mathbf{w} \Vert \leq \dfrac{\sqrt{M \delta}.\Vert \mathbf{w} \Vert}{\min\lbrace \Vert \mathbf{a}_{1}\Vert,...,\Vert \mathbf{a}_{M}\Vert \rbrace}  
\end{equation}
By using the assumption of the lemma,

\begin{equation}  \Vert \mathbf{B}^\mathtt{H}\mathbf{w} \Vert  < \dfrac{\sqrt{M \delta}.\dfrac{ \min\lbrace \Vert \mathbf{a}_{1}\Vert,...,\Vert \mathbf{a}_{M}\Vert \rbrace}{2\sqrt{M \delta}}}{\min\lbrace \Vert \mathbf{a}_{1}\Vert,...,\Vert \mathbf{a}_{M}\Vert \rbrace}
\end{equation}

\begin{equation} \Longrightarrow \Vert \mathbf{B}^\mathtt{H}\mathbf{w} \Vert < \frac{1}{2} .
\end{equation}
\end{proof}

\begin{lemma}
Consider $ \mathbf{B}=[\mathbf{b}_{1}...\mathbf{b}_{M}] $ as a reduced basis (LLL) \cite{LLL82} and $ d_{\mathbf{H}} $ as the minimum distance of the lattice generated by $ \mathbf{H}$, respectively. Then, there is a constant number $ c_{M} $ (independent of $ \mathbf{H} $) such that the LLL-aided decoding method correctly decodes the transmitted signal, if the magnitude of the noise vector is less than $ c_{M} d_{\mathbf{H}} $.
\end{lemma}

\begin{proof} For an LLL reduction, 
\begin{equation} \sqrt{\delta} \leq 2^{M(M-1)}. 
\end{equation}
Therefore, if we consider $c_{M}=\dfrac{2^{-1-M(M-1)}}{\sqrt{M}}$,

\begin{equation} \Vert \mathbf{w} \Vert \leq c_{M} d_{\mathbf{H}} 
\Longrightarrow \Vert \mathbf{w} \Vert \leq \dfrac{1}{2\sqrt{M \delta}}  d_{\mathbf{H}}
\end{equation}
The basis $ \mathbf{B}$ can be written as $ \mathbf{B}=\mathbf{H}^{-\mathtt{H} }\mathbf{U}$ for some unimodular matrix $\mathbf{U}$:

\begin{equation}
\mathbf{B}^{-\mathtt{H} } = (\mathbf{H}^{-\mathtt{H} }\mathbf{U})^{-\mathtt{H} }=\mathbf{H}\mathbf{U}^{-\mathtt{H} }
\end{equation}
Thus, $ \mathbf{B}^{-\mathtt{H} }=[\mathbf{a}_{1},...,
\mathbf{a}_{M}] $ is another basis for the lattice generated by $\mathbf{H}$. Therefore, $\mathbf{a}_{1},...,
\mathbf{a}_{M}$ are vectors from the lattice generated by $\mathbf{H}$, and therefore, the length of each of them is at least $d_{\mathbf{H}}$. Therefore,

\begin{equation}  \Vert \mathbf{w} \Vert \leq \dfrac{1}{2\sqrt{M \delta}}  d_{\mathbf{H}} \leq \dfrac{1}{2\sqrt{M \delta}}  \min\lbrace \Vert \mathbf{a}_{1}\Vert,...,\Vert \mathbf{a}_{M}\Vert \rbrace. 
\end{equation}
Thus, according to lemma 2, LLL-aided decoding method correctly decodes the transmitted signal.
\end{proof}

\begin{lemma} [see \cite{TechRep}]
Assume that the entries of the $ N\times M $ matrix $\mathbf{H}$ has
independent complex Gaussian distribution with zero mean and unit
variance and consider $ d_{\mathbf{H}} $ as the minimum distance of the
lattice generated by $ \mathbf{H} $. Then, there is a constant $\beta_{N,M} $
such that,
\begin{eqnarray}
\Pr\left\lbrace d_{\mathbf{H}}\leq \varepsilon \right\rbrace \leq \left \lbrace
\begin{array}{c}
 \beta_{N,M} \varepsilon^{2N}   \;\;\;\;\;\;\;\;\;\;\;\;\;\;\;\;\;\;\;\;\;\;\;\;\;\;\;\;\;\;\;\;\;\; \textmd{for}  \; M<N  \\
\beta_{N,N} \varepsilon^{2N}.\max\left\lbrace (-\ln \varepsilon)^{N+1},1\right\rbrace \;\; \textmd{for}
\;M=N
\end{array} \right. .
\end{eqnarray}

 \end{lemma}

\begin{thm}
For a MIMO multi-access system (or a point-to-point MIMO system with the V-BLAST transmission) with $ M $ transmit antennas and $ N $ receive antennas, when we use the LLL lattice-aided-decoding,
\begin{equation} \lim_{\rho\rightarrow \infty} \dfrac{-\log P_{e}}{\log \rho}=N.
\end{equation}
\end{thm}
\begin{proof} When $ \Vert \mathbf{w} \Vert \leq c_{M} d_{\mathbf{H}} $, according to lemma 3, we have no decoding error. Thus, 
\begin{equation} P_{e} \leq  \Pr\left\lbrace \Vert  \mathbf{w} \Vert > c_{M} d_{\mathbf{H}} \right\rbrace 
\end{equation}
$$
= \Pr\lbrace c_{M}^{2} d_{\mathbf{H}}^{2}\leq \dfrac{1}{\rho} \rbrace \cdot \Pr \left\lbrace \Vert  \mathbf{w} \Vert > c_{M} d_{\mathbf{H}} \left\vert c_{M}^{2} d_{\mathbf{H}}^{2}\leq \dfrac{1}{\rho}  \right. \right\rbrace + 
$$
\begin{equation}
\sum_{i=0}^{\infty} \Pr \lbrace \dfrac{2^{i}}{\rho} < c_{M}^{2} d_{\mathbf{H}}^{2}\leq \dfrac{2^{i+1}}{\rho} \rbrace \cdot  \Pr \left\lbrace \Vert  \mathbf{w} \Vert > c_{M} d_{\mathbf{H}} \left\vert \dfrac{2^{i}}{\rho} < c_{M}^{2} d_{\mathbf{H}}^{2}\leq \dfrac{2^{i+1}}{\rho} \right. \right\rbrace 
\end{equation}
$$ \leq \Pr\lbrace c_{M}^{2} d_{\mathbf{H}}^{2}\leq \dfrac{1}{\rho} \rbrace +
$$

  \begin{equation}\label{eq:22}  \sum_{i=0}^{\infty} \Pr \lbrace c_{M}^{2} d_{\mathbf{H}}^{2}\leq \dfrac{2^{i+1}}{\rho} \rbrace \cdot \Pr\left\lbrace \Vert \mathbf{w} \Vert^{2} \geq \frac{2^{i}}{\rho}  \right\rbrace
    \end{equation}
    
    The noise vector has complex Gaussian distribution with variance $ \dfrac{1}{2\rho}$ per each real dimension. Thus, by using the union bound, we can bound the second part of each product term as,
\begin{equation}\label{eq:23} \Pr\left\lbrace \Vert \mathbf{w} \Vert^{2} \geq \frac{\gamma }{\rho}  \right\rbrace  \leq \sum_{i=1}^{2N} \Pr\left\lbrace   \vert w_{i} \vert^{2}  \geq \frac{\gamma }{2N\rho}  \right\rbrace   \leq  2N Q \left( \sqrt{\frac{\gamma}{N}} \right)   \leq  2N e^{-\frac{\gamma }{2N}}\end{equation}
Also, for the first part of the product terms, we have,
\begin{equation} \label{eq:24}  \Pr \left\lbrace c_{M}^{2} d_{\mathbf{H}}^{2}\leq \dfrac{\theta}{\rho} \right\rbrace  =  \Pr \left\lbrace  d_{\mathbf{H}}\leq \sqrt{\dfrac{\theta}{c_{M}^{2}\rho}} \right\rbrace \end{equation}
By using (\ref{eq:23}) and (\ref{eq:24}), we can bound (\ref{eq:22}).

\textbf{Case 1}, $M<N$:  
\begin{equation} (\ref{eq:22})\leq \beta_{N,M}\left( \frac{1}{c_{M}^{2}\rho}\right)^{N}  
+  \sum_{i=0}^{\infty}\beta_{N,M} \left( \frac{2^{i+1}}{c_{M}^{2}\rho}\right)^{N} \cdot2N \cdot e^{- \frac{2^{i}}{2N}}
    \end{equation}
  
\begin{equation} =\frac{ \beta_{N,M}}{\rho^{N}}\left( \left( \frac{1}{c_{M}^{2}}\right)^{N}  
+  \sum_{i=0}^{\infty} \left( \frac{2^{i+1}}{c_{M}^{2}}\right)^{N} \cdot2N \cdot e^{- \frac{2^{i}}{2N}}\right) 
    \end{equation} 
 
\begin{equation} \Longrightarrow  P_{e}\leq  \dfrac{c}{\rho^{N}}
\end{equation}
where $ c $ is a constant\footnote{The terms of this series have double exponential parts which ensure its convergence (according to the ratio test).}. Therefore,

\begin{equation}\lim_{\rho\rightarrow \infty} \dfrac{-\log P_{e}}{\log \rho} \geq N. 
\end{equation}

\textbf{Case 2}, $M=N$: 

$$
 (\ref{eq:22})\leq  \beta_{N,N}\left( \frac{1}{c_{M}^{2}\rho}\right)^{N} \max\left\lbrace \left( \frac{1}{2}\ln c_{M}^{2}\rho \right)^{N+1},1\right\rbrace +  
$$ 

\begin{equation} 
\sum_{i=0}^{\infty} \beta_{N,N} \left( \frac{2^{i+1}}{c_{M}^{2}\rho}\right)^{N}\max\left\lbrace \left( \frac{1}{2}\ln \frac{c_{M}^{2}\rho} {2^{i+1}}\right)^{N+1},1\right\rbrace \cdot 2N \cdot e^{- \frac{2^{i}}{2N}} 
    \end{equation}  
We are interested in the large values of $\rho$. For $\rho > c_{M}^2$ and $\ln \rho > 1$,
 \begin{equation}
  (\ref{eq:22}) \leq \beta_{N,N}\left( \frac{1}{c_{M}^{2}\rho}\right)^{N}  (\ln\rho) ^{N+1}
+ \sum_{i=0}^{\infty} \beta_{N,N}\left( \frac{2^{i+1}}{c_{M}^{2}\rho}\right)^{N}(\ln\rho)^{N+1} \cdot 2N \cdot e^{- \frac{2^{i}}{2N}}
  \end{equation}
 
  \begin{equation}
 = \frac{ \beta_{N,N}(\ln\rho) ^{N+1}}{\rho^{N}}\left( \left( \frac{1}{c_{M}^{2}}\right)^{N}  
+ \sum_{i=0}^{\infty} \left( \frac{2^{i+1}}{c_{M}^{2}}\right)^{N} \cdot 2N \cdot e^{- \frac{2^{i}}{2N}}\right) 
  \end{equation}
  
\begin{equation} \Longrightarrow  P_{e}\leq  \dfrac{c^{\prime}\left( \ln \rho\right) ^{N+1}}{\rho^{N}} 
\end{equation}
where $ c^{\prime} $ is a constant. Therefore,

\begin{equation} \lim_{\rho\rightarrow \infty} \dfrac{-\log P_{e}}{\log \rho} \geq \lim_{\rho\rightarrow \infty} \dfrac{\log \rho^{N} -(N+1)\log \left( \ln \rho\right)-\log c^{\prime} }{\log \rho} = N. 
\end{equation}

\end{proof}

In the above proof, we have considered the LLL-aided decoding type I. In this case, the effective noise vector is equal to $\mathbf{w}^{\prime}= \mathbf{B}^{\mathtt{H} }\mathbf{w}$, compared to $\mathbf{w}^{\prime}= \mathbf{H}^{-1}\mathbf{w}$ in zero-forcing. In the previous works \cite{fischer2003} \cite{Wornell} \cite{Mow}, the LLL-aided decoding type II has been used. For the type II method, the effective noise vector is equal to $\mathbf{w}^{\prime}={\mathbf{H}_{red}}^{-1}\mathbf{w} $ and the average energy of its $i$th component is  proportional to the square norm of the $i$th column of ${\mathbf{H}_{red}}^{-\mathtt{H} }$. By using inequality (\ref{eq:lemma1-2}) from lemma 1 (to bound the square norm of the columns of ${\mathbf{H}_{red}}^{-\mathtt{H} }$) and using a similar proof as lemma 2, we can show that the results of lemma 2 and theorem 2 are still valid. Therefore, both of these LLL-aided decoding methods achieve the receive diversity in V-BLAST MIMO systems (or multiple access MIMO systems). However, it is worth noting that the first method is a more natural approach to reduce the power of the entries of the effective noise vector, and has a better performance (see figure 1). For the case of real lattices, a latice-reduction-aided approach similar to type I is recently studied in \cite{Ling2006} and based on the concept of proximity factor, another justification for its superior performance over type II is presented. 

\begin{figure}
  \centering
  \includegraphics[scale=.8,clip]{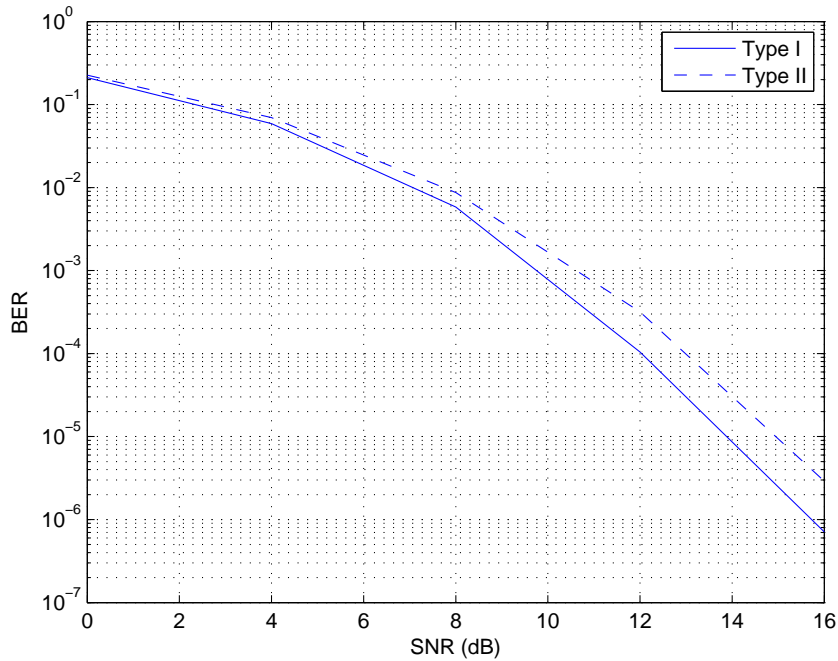}

  \caption{Bit Error Rate of the two LLL-aided decoding methods for $M=6$ transmit antennas and $N=6$ receive antennas with the rate $R=12$ bits per channel use.}
  \label{fig:reduction}
\end{figure}

\section{Relation with the naive lattice-decoding}

When we have a finite constellation, for each pair of constellation points, the pair-wise error probability can be bounded by Chernoff bound (similar to \cite{Tarokh98}). By using the union bound, we can show that the exact ML decoding achieves the diversity order of $N$, the number of receive antennas. However, when we use lattice decoding for a finite constellation and consider the out-of-region decoded lattice points as errors, achieving the maximum diversity by lattice decoding is not trivial anymore. Nonetheless, by using lemma 4, we can show that this suboptimum method (called the naive lattice decoding \cite{DamenElgamalCaire}) still achieves the maximum diversity.

\begin{thm}
For a MIMO multi-access system (or a point-to-point MIMO system with the V-BLAST transmission method) with $ M $ transmit antennas and $ N $  receive antennas, when we use the naive lattice decoding,

\begin{equation} \lim_{\rho\rightarrow \infty} \dfrac{-\log P_{e}}{\log \rho}=N.
\end{equation}
\end{thm}

\begin{proof}
When $ \Vert \mathbf{w} \Vert \leq \frac{1}{2} d_{\mathbf{H}} $, we have no decoding error. Thus, by using $\frac{1}{2} $ instead of $c_{M} $ in the proof of theorem 2, we can bound $P_{e} $ by bounding $  \Pr\left\lbrace \Vert  \mathbf{w} \Vert > \frac{1}{2} d_{\mathbf{H}} \right\rbrace $. Therefore, we can obtain the same result as theorem 2.
\end{proof}

In \cite{DamenElgamalCaire}, it is shown that for the naive lattice decoding, we can find a family of lattices (generating a family of space-time codes) which achieves diversity order of $M$ ($ M\leq N$ is the number of transmit antennas). The current result shows that even if we use the codes generated by the integer lattice, the naive lattice decoding achieves the maximum receive diversity of $N$ (number of receive antennas). 

\section{Conclusions}

We have shown that LLL-aided zero-forcing, which is a polynomial-time algorithm, achieves the maximum receive diversity in MIMO systems. By using LLL reduction before zero-forcing, the complexity of the MIMO decoding is equal to the complexity of the zero-forcing method with just an additional polynomial-time preprocessing for the whole fading block. Also, it is shown that by using the naive lattice decoding, instead of ML decoding, we do not loose the receive diversity order.

\section{Acknowledgment}

The authors would like to thank M. O. Damen for helpful discussions.

\end{document}